\documentclass[PAPER,cernpreprint, atlasdraft=false, UKenglish, texmf, orcidlogo]{atlasdoc}
\usepackage{atlaspackage}
\usepackage{atlasbiblatex}

\usepackage{atlasphysics}

\usepackage{comment}
\usepackage{float}
\graphicspath{{logos/}{figures/}}

\usepackage{ANA-HION-2023-13-PAPER-defs}

\AtlasTitle{Measurement of coincident photon-initiated processes in ultra-peripheral Pb+Pb collisions with the ATLAS detector}

\AtlasAbstract{%
The Lorentz-contracted  electromagnetic fields of the ions in ultra-relativistic heavy-ion collisions
generate intense quasi-real photon fluxes.
These lead to photon-induced interactions that are observed in
ultra-peripheral collisions (UPCs), such as vector meson and lepton-pair production.
The high photon flux also enables the occurrence of multiple photon-induced
processes in a single collision.
Presented is the first measurement of the coincident production of
$\gamma\gamma \rightarrow \mu^{+}\mu^{-}$ and $\gamma+A\rightarrow\rho^{0}+A$ in
UPC Pb+Pb collisions at centre-of-mass energies of 5.02~TeV and 5.36~TeV with
the ATLAS detector at the Large Hadron Collider.
The coincident $\rho^{0}$ is identified via its decay into a $\pi^+\pi^-$ pair,
with the selected $\pi^+\pi^-$ sample also containing the
non-resonant continuum contribution.
The rate of the coincident process relative to the exclusive
$\gamma\gamma \rightarrow \mu^{+}\mu^{-}$ process is measured differentially
in intervals of forward event activity, quantified by the Zero Degree Calorimeters.
The relative rate of coincident $\rho^{0}$ production, including non-resonant $\pi^+\pi^-$ contributions,
is measured to be $(9.3\,\pm0.4\,\mathrm{(stat.)}\,\pm0.2\,\mathrm{(syst.)})\times10^{-3}$,
summed over forward event activity.
Correlations between the dimuon kinematic properties, such as its mass,
and the coincident $\rho^{0}$ meson production rate, are also presented.
These measurements confirm the presence of multi photon-induced processes in UPC collisions,
and can provide new insight into the impact parameter dependence of
photon-induced vector meson production.
}

\AtlasRefCode{HION-2023-13}

\PreprintIdNumber{CERN-EP-2025-071}

\arXivId{2504.07795}

\HepDataRecord{2910759}

\AtlasJournalRef{\PLB 881 (2026) 140909}
\AtlasDOI{10.1016/j.physletb.2026.140909}

\hypersetup{pdftitle={ATLAS document},pdfauthor={The ATLAS Collaboration}}

\begin{document}

\maketitle

\section{Introduction\label{sec:intro}}

Ultra-peripheral heavy-ion collisions (UPCs) provide a unique
environment to study photon-initiated scattering processes at high energy.
In these collisions, the impact parameter is typically larger than
the sum of the radii of the colliding nuclei, so that hadronic interactions between the nuclei are suppressed.
Instead, the intense electromagnetic fields generated by the
relativistic ions dominate the interaction, allowing photon--photon
and photon--nucleus interactions to occur.
Such photon-initiated processes were studied extensively
at the Relativistic Heavy Ion Collider (RHIC)~\cite{RHIC}
and at the Large Hadron Collider (LHC)~\cite{Evans:2008zzb}.
These include cases where both the nuclei emit photons,
leading to processes such as
light-by-light scattering \gamgamgamgam~\cite{HION-2016-05,HION-2018-19,HION-2019-08,CMS-FSQ-16-012,CMS-HIN-21-015},
and dilepton production from photon fusion, \gamgamll~\cite{ALICE:2013wjo,HION-2016-02,HION-2018-11,HION-2020-10,HION-2021-16,CMS-HIN-21-009,ALICE:2023mfc,STAR:2019wlg}.
Another category of UPC processes are those where the photon emitted by one nucleus
fluctuates to a quark--anti-quark pair and
interacts with the second via the strong interaction,
producing dijets~\cite{HION-2022-15} or
exclusive vector mesons, like \rhoz~\cite{STAR:2002caw,STAR:2007elq,STAR:2011wtm,STAR:2017enh,ALICE:2015nbw,ALICE:2020ugp,ALICE:2021jnv,ALICE:2024ife,CMS-FSQ-16-007,LHCb:2025fzk}
or $J/\psi$~\cite{ALICE:2012yye,ALICE:2014eof,ALICE:2018oyo,ALICE:2019tqa,ALICE:2021gpt,ALICE:2023jgu,ALICE:2023svb,CMS-HIN-12-009}.
At higher masses, these processes are sensitive to the nuclear
parton distribution functions (nPDFs), especially the gluon nPDFs,
while at lower vector meson masses, the processes probe the
coherent interactions of the soft pomeron ($\mathrm{I\!P}$) with the nucleus.

There has long been theoretical interest in UPC processes that are
higher order in the atomic number ($Z$) due to the multiple emission of photons from one
or both of the nuclei~\cite{Bertulani:1987tz,Klein:1999qj,Baur:2001jj,vanHameren:2017krz,Harland-Lang:2021ysd,Azevedo:2023vsz}.
For example, processes producing multiple pairs in the same
collision are necessary for the preservation of unitarity in
theoretical calculations of lepton-pair production at high field
strength~\cite{Baur:2001jj,Harland-Lang:2021ysd,}.
Separately, it has also been suggested that multiple lepton-pair production could
affect measurements of exclusive UPC processes in heavy-ion collisions, by reducing the
survival factor of the exclusive final
state~\cite{Harland-Lang:2021ysd,Hencken:2006ir}.
While theoretical calculations
have often focused on multiple lepton-pair production, the atomic
mass number ($A$) of the Pb nucleus is  large enough that
contributions to multi-photon processes from coherent diffraction
should not be neglected.
Indeed, the probability
for \rhoz\ production via $\gamma+\mathrm{ I\!P}$ scattering in a
typical UPC \PbPb\ event can be of $\mathcal{O}(1\%)$~\cite{Klein:1999qj}
once the impact parameter has been constrained by the requirement that
another UPC process has occurred.
Thus, any UPC scattering process, including lepton-pair
production, may have non-negligible probability for ``coincident'' -- in the same \PbPb\ collisions --
$\gamma+\mathrm{I\!P}\rightarrow \rho$ production.

The kinematics of \gamgamll\ pair production are sensitive to
the impact-parameter dependence of nuclear photon
energy distributions.
In particular, the fluxes for the photons in the \gamgamll\ processes
fall rapidly with increasing transverse distance from the nucleus,
and the dependence is stronger for higher photon energies~\cite{Klein:2020fmr}.
Consequently, selections on the lepton-pair kinematics, such as the
invariant mass of the lepton pair, can be used to intentionally
bias the impact-parameter distribution, so that measurements of
the coincident $\rhoz$ yield can be used to test the understanding
of the impact-parameter dependence of $\gamma+\mathrm{ I\!P} \rightarrow \mathrm{V}$
processes.
Thus, studies of coincident $\gamma\gamma$ production of leptons and
$\gamma+\mathrm{ I\!P}$ production of $\rhoz$ mesons not only provide
a novel test of theoretical predictions of photon-induced processes
at high electromagnetic field strength, but will also shed new light
on exclusive QCD processes at high gluon field strength.

The relative rate is presented for the coincident occurrence of \gamgammumu\ and
a \rhoz\ in UPC \PbPb\ collisions.
The rate is calculated relative to the exclusive
\gamgammumu\ process, where ``exclusive'' refers to
the absence of additional particles in the event, except for the two muons.
The measurement is performed with the ATLAS detector using LHC data recorded
at centre-of-mass energies per nucleon-pair (\snn) of 5.02~\TeV\ and 5.36~\TeV,
with integrated luminosities of 1.44~\inb\ and 3.17~\inb, respectively.
The 5.02~\TeV\ data were recorded in 2018 and the 5.36~\TeV\ data were
recorded in 2023 and 2024.
The \gamgammumu\ process, which is studied extensively in
UPC events in ATLAS~\cite{HION-2016-02,HION-2018-11,HION-2020-10},
is used to ``tag'' the events, which are then used to search
for the presence of a single additional \rhoz,
with the requirement that no additional tracks are present in the event.
The \rhoz\ is identified via its decay into a pair of oppositely charged particle tracks whose invariant mass,
assuming that the tracks are charged pions, is required to lie in the range 300--1500~\MeV\ and to be consistent with the
\rhoz\ mass distribution obtained from the \starlight~\cite{Klein:2016yzr} event generator
after being processed through a full \GEANT\ simulation~\cite{Agostinelli:2002hh}
of the ATLAS detector~\cite{SOFT-2010-01}.
The pion pair production via \rhopipi\ decay interferes with direct~\pipi\ production,
where a photon fluctuates directly to a pair of charged pions~\cite{Klein:2016yzr}.
The measurements presented here include the rates for both the processes.
The relative rate of the coincident \rhoz\ production
is evaluated in intervals of the invariant mass of the
\mumu\ pair, its rapidity, and the energy of the higher energy
photon in the tagging \gamgammumu\ process.
The coincident \rhoz\ production is also studied as a function of the
forward neutron activity in the events,
which can provide additional qualitative constraints on the
impact parameter of the collision.


%

%
\section{ATLAS detector\label{sec:detector}}

The ATLAS experiment~\cite{PERF-2007-01,GENR-2019-02} at the LHC is a multipurpose particle detector
with a forward--backward symmetric cylindrical geometry and a
near \(4\pi\) coverage in solid angle\footnote{\AtlasCoordFootnote}.
It consists of an inner tracking detector (ID) surrounded by a thin superconducting solenoid
providing a \SI{2}{\tesla} axial magnetic field, electromagnetic and hadronic calorimeters, and a muon spectrometer.
The ID provides charged-particle tracking in the range of \(|\eta| < 2.5\).
The high-granularity silicon pixel detector covers the interaction
region and typically provides four measurements (hits) per track,
the first hit generally being in the insertable B-layer (IBL). %
It is followed by the SemiConductor Tracker (SCT), which usually provides eight measurements per track.
These silicon detectors are complemented by the transition radiation tracker,
which enables radially extended track reconstruction up to \(|\eta| = 2.0\).
The calorimeter system consists of a liquid-argon (LAr) electromagnetic
calorimeter covering $|\eta|<3.2$, a steel/scintillator sampling
hadronic calorimeter covering $|\eta|<1.7$, a LAr hadronic calorimeter
covering $1.5 < |\eta| < 3.2$, and a LAr forward calorimeter (FCal)
covering $3.1 < |\eta| <4.9$.
The muon spectrometer surrounds the calorimeters and is based on
three large superconducting air-core toroidal magnets with eight coils each.
The field integral of the toroids ranges between \num{2.0} and \qty{6.0}{\tesla\metre}
across most of the detector.
The muon spectrometer includes a system of precision tracking chambers up to
\(|\eta| = 2.7\) and fast detectors for triggering up to \(|\eta| = 2.4\).
Two Zero Degree Calorimeters (ZDCs), which measure neutrons emitted at
small rapidity separation from the incident nuclei, are used
for offline event selection.
The ZDCs are located symmetrically at a distance of $\pm140$~m
from the nominal IP %
and cover $|\eta|>$8.3 along the beam axis.
Each ZDC consists of four modules, each containing slightly
more than one interaction length of tungsten absorber.
A two-level trigger system~\cite{TRIG-2016-01,TRIG-2022-01}, the first (L1) implemented in hardware and
the second (HLT) implemented in software, is used to select events
for this measurement.
An extensive software suite~\cite{SOFT-2022-02} is used in the
reconstruction and analysis of real and simulated data, in detector
operations, and in the trigger and data acquisition systems of the experiment.


%

%
\section{Analysis procedure\label{sec:analysis}}

Events are selected using a muon trigger that requires
the total transverse-energy measured in the calorimeters
at L1, $E_{\mathrm{T}}^{\mathrm{L1,calo}}$, to be less than 50~\GeV\ and the presence of at least one muon
having a transverse momentum (\pt) greater than $4~\GeV$
at L1 and at the HLT\@.
Performance of this muon trigger, but without the $E_{\mathrm{T}}^{\mathrm{L1,calo}}<50~\GeV$
requirement, is summarized in Ref.~\cite{TRIG-2018-01}.
Events used in the analysis are required to be recorded during stable
running conditions of the LHC, to have no detector hardware or readout
error, and to have a reconstructed collision vertex~\cite{DAPR-2018-01,ATL-PHYS-PUB-2015-026,ATL-PHYS-PUB-2019-015}.
Charged-particle tracks and collision vertices are reconstructed from
hits in the ID using standard methods~\cite{PERF-2015-08}.
In addition to  standard charged-particle tracks reconstructed using the whole ID,
hereafter called ID-tracks, the measurement also uses tracks
reconstructed using hit information in only the pixel detector~\cite{ATL-PHYS-PUB-2022-033},
hereafter called pixel-tracks.
Muons are reconstructed by combining ID-tracks with tracks
reconstructed in the muon spectrometer.
As stated before, the ZDCs measure neutral particles emitted at
small angles relative to the beam direction.
Increased neutron emission is associated, on average, with smaller
impact parameters and therefore with different impact-parameter-dependent
photon-flux and photon-energy distributions.
The analysis is performed in the following selections
of the ZDC activity:
\begin{itemize}
\item \OnOn : no neutrons detected in either arm of the ZDC\@.
These events correspond to large impact parameters, where the nuclei remain entirely intact.
\item \OnYn : at least one neutron detected in one arm of the ZDC and no neutrons in the other arm.
These events correspond to slightly smaller impact parameters and the dissociation of one nucleus.
\item \YnYn : at least one neutron detected in each arm of the ZDC\@.
These events correspond to even smaller impact parameters and the breakup of both the nuclei.
\end{itemize}
Two combined intervals of the ZDC activity are also studied:
\begin{itemize}
\item \OnXn : no neutrons detected in one arm of the ZDC with no requirements on the other arm,
\item \XnXn : no requirements on either of the ZDC arms.
\end{itemize}
The migration between the described ZDC selections due to electromagnetic (EM)
``pile-up,'' where an independent UPC event
-- involving two different Pb nuclei -- occurs at the same time as an event of interest
is corrected for using the procedure detailed in Ref.~\cite{HION-2016-02}.
The correction procedure uses a migration matrix
whose entries correspond to migration probabilities
between the \OnOn, \OnYn, and \YnYn\ configurations due to pile-up.
The calculation of these probabilities are described in Ref.~\cite{HION-2016-02},
and depend on the cross-sections of inclusive single- and double-dissociative
processes and the instantaneous luminosity during data taking.

Following Ref.~\cite{HION-2020-10}, for selecting the \gamgammumu\ process, events are required to have
two reconstructed offline muons with $|\eta|<2.4$ and $\pt>3.7$~\GeV.
At least one of the two muons in the pair is required to match
the single-muon trigger that selected the event.
The muons are required to have opposite charges and
satisfy the \textit{medium}
muon selection requirements~\cite{PERF-2015-10}.
Both the muons are required to have a transverse impact parameter
relative to the beam-line, $|d_{0}|$, of less than 1~mm
and to have a longitudinal  impact parameter relative to the primary-vertex,
$|z_0\sin(\theta)|$, of less than 1~mm.
Additional selections are required on the following quantities
(indices `$a$' and `$b$' refer to the two muons in the pair):
\begin{itemize}
\item \textit{momentum asymmetry } $\Asym\equiv|\pT^{a}-\pT^{b}|/(\pT^{a}+\pT^{b})$:
a measure of the \pt\ mismatch between the two muons,
\item $\ptbar\equiv(\pT^{a}+\pT^{b})/2$:
the average \pt\ of the two muons in the pair,
\item \textit{acoplanarity} $\Aco\equiv(1-|\phi^{a}-\phi^{b}|/\pi)$:
a measure of how back-to-back the muons are in the transverse plane,
\item $\kperp\equiv\pi\Aco\ptbar$: the transverse momentum of one muon perpendicular to the \pt\ axis of the other.
\end{itemize}
Dimuons from the coherent \gamgammumu\ process are typically back-to-back
in $\phi$ with nearly matched \pt, so they usually have
small \Aco, \Asym, and \kperp\ (see Ref.~\cite{HION-2020-10,Baltz:2009jk,Bertulani:2005ru} and references therein).
Therefore the muon pairs are required to have $\Asym<0.06$ and $\kperp <150$~\MeV.
The muon pairs are also required to have $\ptbar>4~\GeV$  to match
the requirements applied in the trigger.
With these selections, the muon pairs observed in UPC events
mostly originate from the \gamgammumu\ process~\cite{HION-2020-10}.
The total number of exclusive \gamgammumu\ pairs satisfying these requirements
in the \OnOn, \OnYn, and \YnYn\ ZDC selections are $1.05\times10^{5}$,
$4.8\times10^{4}$, and $1.3\times10^{4}$, respectively.
Contributions from dissociative processes are estimated via the procedure
used in a prior ATLAS measurement~\cite{HION-2016-02},
by fitting the dimuon acoplanarity distributions using a
two component-fit of the form:
\begin{eqnarray}
\frac{1}{N}\frac{dN}{d\Aco}= (1-\fdis)\frac{1}{N}\frac{dN_{\mathrm{EPA}}}{d\Aco} + \fdis\frac{1}{N}\frac{dN_{\mathrm{dis}}}{d\Aco},
\label{eq:fdis}
\end{eqnarray}
where the template for the dissociative process,
$\frac{1}{N}\frac{dN_{\mathrm{dis}}}{d\Aco}$, is obtained from
the \lpair\ event generator~\cite{Vermaseren:1982cz}, and the
template for coherent dimuons, $\frac{1}{N}\frac{dN_{\mathrm{EPA}}}{d\Aco}$,
is obtained from \starlight.
Due to the $\kperp<150~\MeV$ requirement, the muons in the
pair are almost back-to-back in $\phi$, and the fraction of muons
from dissociative processes is strongly suppressed.
The residual contamination is negligible for the \OnOn\ ZDC selection
and is estimated to be less than 1\%, 2\%, 3\% and 1\%, for the
\OnXn, \OnYn, \YnYn, and \XnXn\ ZDC selections, respectively.
While the $\Aco$, $\Asym$, $\kperp$ and \ptbar\ quantities are defined
above in the context of muons,
identical definitions are also used for candidate \rhopipi\ track pairs
with the labels `$a$' and `$b$' indicating the two pion track candidates.

The signal selection for this measurement requires \gamgammumu-tagged events
containing exactly two additional tracks having opposite charge.
For the candidate \rhopipi\ products, the ID-tracks are required to have at least six silicon (pixel+SCT) hits,
less than or equal to two silicon holes, and at most one pixel hole,
where a hole is defined as the absence of a hit predicted by the track trajectory
in between the innermost and outermost measurement.
The pixel-tracks are required to have at least three hits and no holes.
Both the ID-tracks and pixel-tracks are required to have $|d_{0}|<1.5$~mm
and $|z_0\sin(\theta)|<1.5$~mm.
The sets of ID-tracks and pixel-tracks are merged with removal of duplicates by
requiring that only pixel-tracks with a
$\sqrt{\Delta\eta^2+\Delta\phi^2}$ separation of more than
0.05 from all ID-tracks in an event are used.
The two pion track candidates are required to have $\pt>100$~\MeV\ and $|\eta|<2.5$,
which defines the fiducial region of the analysis.

As specified earlier, the measurement is done at two different
centre-of-mass energies.
A \starlight\ simulation shows that the
cross-section for independent \rhoz\ production changes only by $\sim2$\% between
the two collision energies, while for the \gamgammumu\ process
with the previously described kinematic requirements on the muons,
the change in the cross-section is $\sim8$\%.
Due to this relatively small change in the expected cross-sections,
and also to gain statistical precision, which is limited when separately
measuring for the two energies, the candidate processes from
the two energies are combined into a single measurement.
As a check, the data for the two energies were analyzed separately
and were found to be consistent within statistical uncertainties.
The measurement of the {\em relative} \rhopipi\ rates also implies cancellation of the efficiency for
triggering on and reconstructing the \gamgammumu\ process.
Thus, no dimuon trigger or reconstruction efficiency corrections are applied
in the analysis.

To reduce backgrounds in the sample of \rhopipi\ candidates,
additional selections of $\Aco<0.15$ and $\Asym<0.2$ are imposed
on the track pairs from the \rhopipi\ candidate process.
These selections are made because
coherently produced \rhoz\ mesons are expected to have small \pt,
and consequently the pions from their decays are expected
to be nearly back-to-back in $\phi$ and have almost no \pt\ mismatch.
These selections are found to reject $\sim80\%$ of track-pairs
in events that have three additional tracks besides
the tagging dimuon.
In the literature, coherently produced \rhoz\ candidates are typically selected by applying
\pt\ based selections on the \rhoz\ candidates (see for example Ref.~\cite{ALICE:2020ugp}).
Here, an acoplanarity- and asymmetry-based selection is used as it was found to improve
the performance in rejecting background pairs from three-track events.
A cross-check is also done by selecting \rhoz\ candidates by requiring
the \pt\ of the candidate to be less than 200~\MeV, instead of the
$\Aco<0.15$ and $\Asym<0.2$ based selection.
The results from the \pt-based selection are consistent with the \Aco-
and \Asym-based selections.
For the \rhoz\ candidate tracks, distributions of the
pair mass (\mpipi), \Aco, \ptbar, and \Asym\ are studied to assess
residual backgrounds.
These backgrounds can arise from
other UPC processes, such as another coincident \gamgammumu\
or \gamgamee\ process, from very low multiplicity (UPC) jets,
and from very peripheral hadronic \PbPb\ interactions.
To obtain the \rhopipi\ yields, the \mpipi\ distributions are fitted with Monte Carlo (MC)
templates obtained from events simulated using the \starlight\
event generator, which are then processed
through a full \GEANT\ simulation
of the ATLAS detector.
The signal templates used in the fit are obtained from the MC simulation of
the \rhopipi\ process, including direct \pipi\ production.
The background templates are formed from simulated events for
the \gamgamee\ and \gamgammumu\ processes.
The reconstructed tracks in these MC samples are required to
satisfy the same track selections used in the data analysis.
An additional background template composed of pairs of tracks
in data events with multiplicity between five and ten tracks
(excluding the tagging dimuon) is also included in the fits.
This latter template accounts for possible backgrounds from
low-multiplicity jets, and very peripheral hadronic
Pb+Pb interactions.

\begin{figure}[ht!]
\centering
\includegraphics[width=1.0\textwidth]{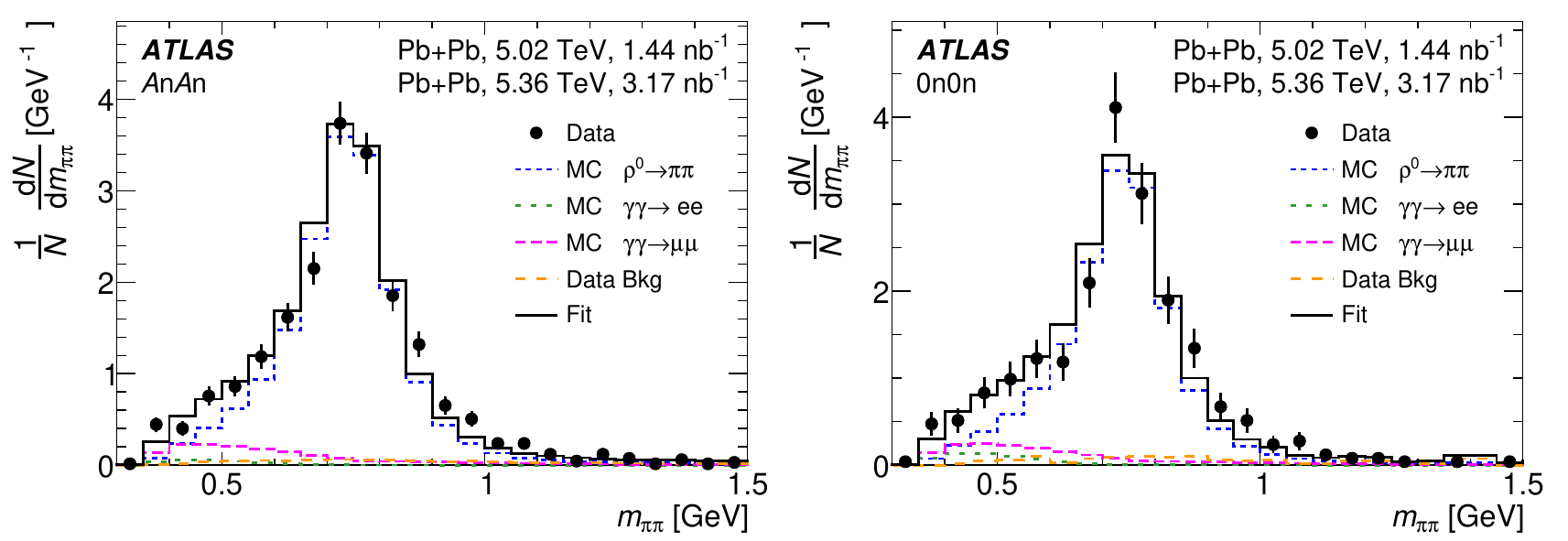}
\caption{
Template fits to the \mpipi\ distributions for the \rhoz\ candidates.
The two panels show the data distributions, the template fit, and the contributions
from the different templates to the fit
for the (left) \XnXn\ and (right) \OnOn\ ZDC selection.
\label{fig:template_fits}
}
\end{figure}

\begin{figure}[ht!]
\centering
\includegraphics[width=1.0\textwidth]{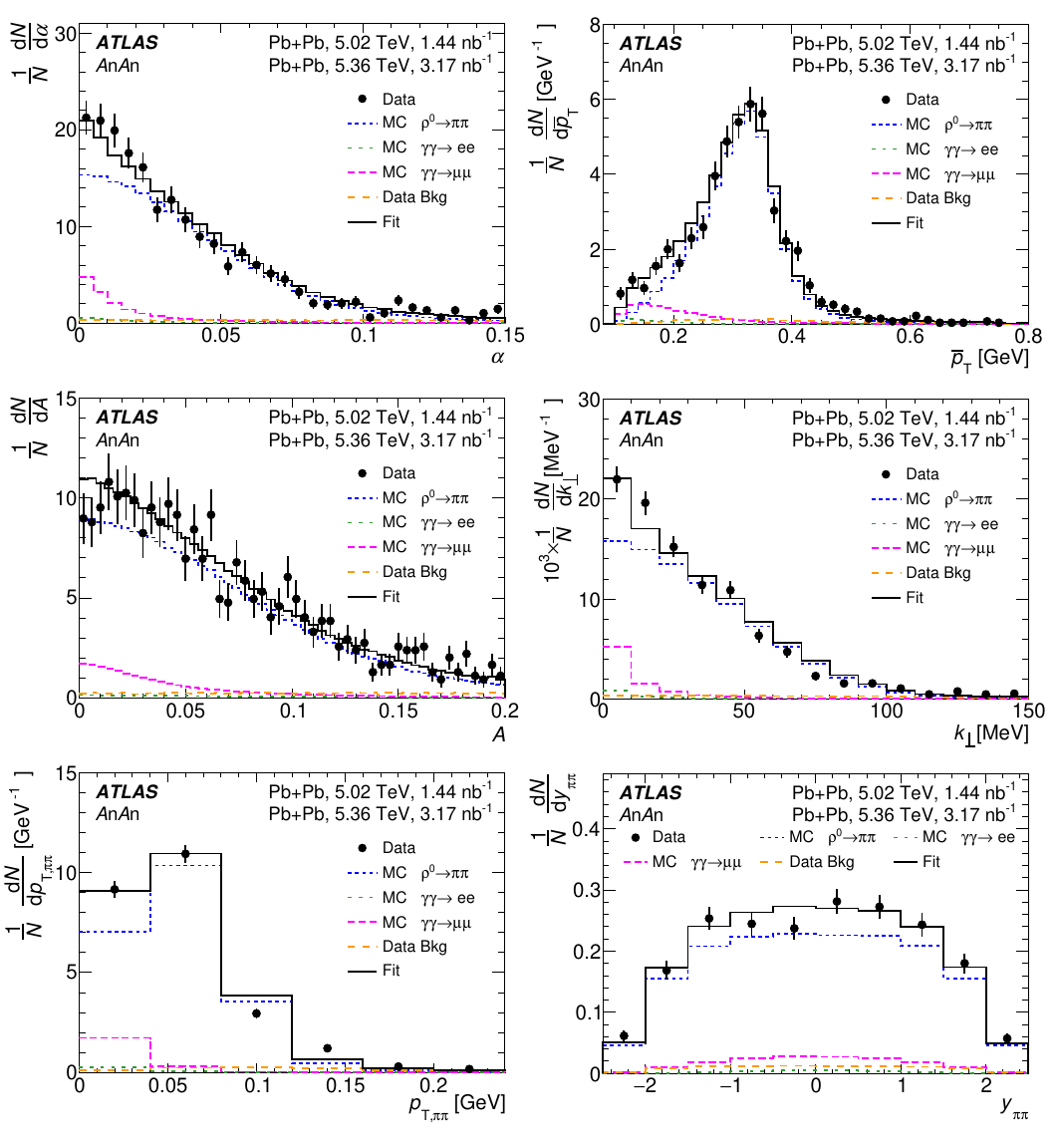}
\caption{
Distributions of track-pair quantities calculated for the \rhopipi\ candidates.
Each panel shows the distribution measured in the data, and contributions
from different MC templates, and their sum.
The template contributions are obtained from fitting the
\mpipi\ data distribution.
The distributions are shown for the \XnXn\ ZDC selection.
\label{fig:template_fits_v2}
}
\end{figure}

Figure~\ref{fig:template_fits} shows the results of template fits
to the \mpipi\ distributions for the \XnXn\ and \OnOn\ ZDC selections,
including the contributions from the signal and the three background templates.
The contribution from the \gamgamee\ background is found to be smaller than
that from the \gamgammumu\ background.
This is due to the reconstruction efficiencies for electron-pairs being smaller
than that for muon-pairs.
Statistical uncertainties in the contributions from the different templates
are evaluated using re-samplings of the measured distributions, assuming
Poisson-distributed statistical uncertainties for each point.
Each of the re-sampled distributions is fitted using the
different templates, and the standard deviation of the resulting
template fractions is taken as the statistical uncertainty.
For the distributions shown in Figure~\ref{fig:template_fits},
the values of the fit-$\chi^2$ divided by the number of degrees of freedom are between 1 and 2.
Distributions of \Aco, \ptbar, \Asym, \kperp, \pTpipi\ and \ypipi\ with template fractions
obtained from the \mpipi\ fits are shown and compared with the data distributions
in Figure~\ref{fig:template_fits_v2} for the \XnXn\ ZDC selection.
The signal counts obtained from the template fits are corrected for inefficiencies
arising from the track reconstruction and the $\Aco<0.15$ and $\Asym<0.2$
requirements imposed on the tracks.
The efficiency correction factor is obtained using the \rhopipi\
MC sample described above and it is defined as the
ratio of the number of events in which the two pion tracks are successfully
reconstructed (i.e., satisfy all offline selections on the tracks and on the track-pair)
to the number of events for which the generated \rhoz\ decayed into two pions
in the measured fiducial region.
The reconstruction efficiency depends on the kinematics of the
two pions from the decay.
For instance, while the efficiency correction has an average value of approximately 1.3,
it varies from $\sim$1.7 at $\ptbar=150~\MeV$
to $\sim1.2$ at $\ptbar=300~\MeV$, beyond which it increases only gradually.
However, as demonstrated in Figure~\ref{fig:template_fits_v2},
the kinematic distributions in the data are well described by the
MC signal templates, so a single average efficiency correction
is applied to the yield obtained from the template fits.

\section{Systematic uncertainties\label{sec:systematics}}
Systematic uncertainties in the \rhoz\ yields are obtained
by varying various aspects of the analysis.
The nominal \rhopipi\ yields are evaluated using template fits
to the \mpipi\ distributions (Figure~\ref{fig:template_fits}).
To test the stability of the results and the ability of the templates
to describe additional features in the data, simultaneous fits are
made to the one-dimensional distributions for \mpipi, \Aco,
\ptbar\ and \Asym, by treating them as independent distributions,
and requiring identical fit fractions for the different templates.
The variation in the result is included as a systematic uncertainty.

The template fits use a data-driven background template
obtained from events having charged-particle multiplicities, excluding the muons, in the range of $[5,10]$.
At the lower end of this range, events can potentially have contributions
from UPC multi-$\rhoz$ production processes~\cite{Klein:2016yzr,ALICE:2024kjy}.
To assess the impact of such events, the template fits are performed with
the data-driven background template restricted to the multiplicity
interval of 7--10, and the variation relative to the nominal result is taken
as a systematic uncertainty. This variation more generally tests the sensitivity of the analysis to
the shape of the background template at lower multiplicity where physical correlations may be more significant.
To check for possible backgrounds from the $\omega\rightarrow\pi^+ \pi^- \pi^0$
process, where only the two charged pions would get reconstructed
as tracks, the analysis was repeated including such an additional
background-template obtained from \starlight.
Including this additional background did not have any
impact on the extracted \rhoz\ yields.
This is because the $\pi^+ \pi^-$ pair from the $\omega\rightarrow\pi^+  \pi^-  \pi^0$
decay are not back-to-back and are rejected by the acoplanarity
selections that are imposed on the $\pi^{+} \pi^{-}$ pair.
To check the stability of the results against alternative track
selection requirements, with their own associated tracking efficiency,
the full analysis is redone using only the pixel-tracks.
The difference between these results and those obtained from the analysis using the combined track
collections is then used as a systematic uncertainty.
This is usually the dominant systematic uncertainty in the measurement,
and is typically between 2\%--8\%.
For the \YnYn\ ZDC selection, where the sample size is limited,
this uncertainty can be much larger and become comparable to
the statistical uncertainty.

A potential background in the measurement could result from UPC pile-up
events in which the \gamgammumu\ pair and the \rhopipi\ pair arise from
independent \PbPb\ UPC events.
This background is suppressed by the strong 1.5~mm vertex-pointing requirements
imposed on the tracks in the longitudinal direction relative to the
primary vertex. The potential background surviving this selection is
estimated by using the measured cross-section for UPC exclusive \rhoz\ production
at the LHC~\cite{ALICE:2020ugp}.
For the luminosity used in this
analysis, the estimated number of pile-up \rhopipi\ pairs having a vertex within $\pm1.5$~mm
of an independent \gamgammumu\ event is $\lesssim 2$ for the combined \XnXn\ ZDC interval.
To check for possible pile-up contamination in a data-driven manner, the $|z_{0}\sin\theta|<1.5$~mm
requirement on the tracks is relaxed to 3~mm.
This change allows additional pile-up-\rhoz\ mesons to be included in the analysis,
and the difference between the results with the two different vertex-pointings
then constitutes the systematic uncertainty resulting from pile-up \rhoz\ contribution.

As described before, the MC simulation is used to obtain both the
reconstruction efficiencies and the templates used in the fits.
Uncertainties in the mis-modelling of the ID material~\cite{PERF-2015-07} are propagated
to the measurement by re-simulating the detector response
with variations in the amount of ID material.
The following variations are considered separately:
the passive material of the ID is scaled up by 5\%,
the passive material of the IBL is scaled up by 10\%,
and the passive material in the services region is scaled up by 25\%.
An additional systematic uncertainty in the modelling of the ATLAS detector
is estimated by varying the ``physics list''~\cite{Geant4PhysicsLists2}
used in the \GEANT\ simulation.
The physics list is changed from the default \verb=FTFP_BERT= to \verb=QGSP_BIC=.
The  changes in the results obtained from the described variations in detector material and
physics list are separately combined with the other systematic uncertainties.
It was stated previously that the analysis does not correct for the trigger and
reconstruction efficiencies of the tagging \gamgammumu\ process,
as the conditional yield of the $\rho$ is not affected by the efficiency
of detecting the tagging process.
However, as a check, the measurement was repeated, while correcting for
the effects of trigger and reconstruction efficiencies of the muon pair.
The results with these corrections applied were consistent with the
measurements without the muon pair efficiency corrections.

As mentioned before, the migration between different ZDC selections
due to EM pile-up is corrected using the procedure described in Ref.~\cite{HION-2016-02}.
The correction is sensitive
to uncertainties in the single and double
EM dissociative cross-sections, the hadronic cross-section
and the uncertainty in the measured luminosity.
The effects of these uncertainties in the pile-up correction  are evaluated separately for $\pm
1\sigma$ changes in each of the associated quantities and the
resulting changes in the \rhopipi\ yields in each ZDC selection are
included in the systematic uncertainties.
These variations do not affect the \XnXn\ selection, where
no ZDC selections are required, or the \OnOn\ selection,
as there is no migration of events into this ZDC selection
and the outflow exactly cancels when taking the ratio
to the number of exclusive \gamgammumu\ events.

The total systematic uncertainty is taken to be the quadrature sum of
all the described contributions.
Overall, except for the ZDC migration uncertainty,
which is $\sim3\%$ for the \OnYn\ and \YnYn\ selections,
and the systematic uncertainty from the ``pixel-track only'' analysis,
the individual systematic uncertainties discussed above are much
smaller than the statistical uncertainties, and are at the
level of $\lesssim1\%$ for the inclusive \mmumu\ and \ymumu\ interval.
%


%

%
\section{Results\label{sec:result}}

\begin{figure}[h]
\centering
\includegraphics[width=1.0\textwidth]{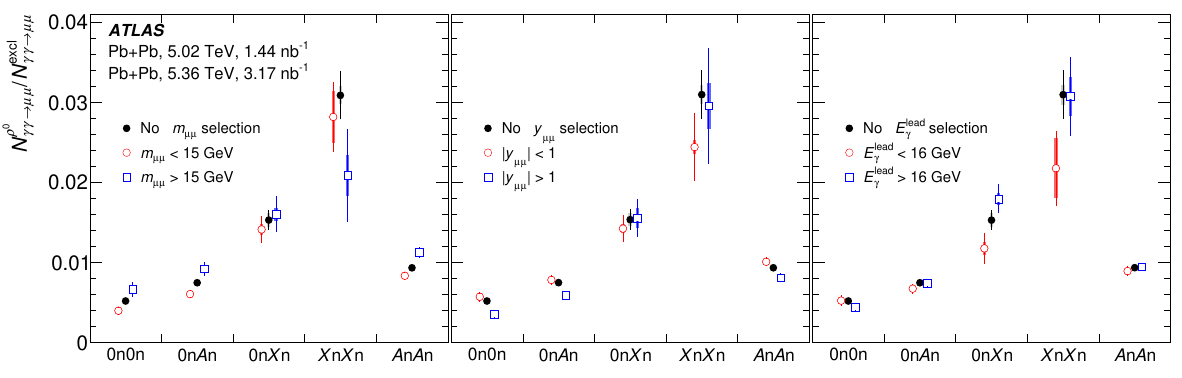}
\caption{
The ratio of the coincident \rhoz\ production in \gamgammumu\ events to
the number of exclusive \gamgammumu\ events for
different ZDC selections.
The left, centre and right panels compare different
\mmumu, \ymumu\ and \egammalead\ (energy of the leading photon)
intervals, respectively.
The error-bars and bands indicate statistical and
systematic uncertainties, respectively.
The coincidence rates are measured in the fiducial acceptance of the
ATLAS detector, namely for $|\eta|<2.5$ and $\pt>100$~\MeV\ for the
pions from the \rhoz\ decay.
The rates are corrected for the detector reconstruction efficiency,
while no correction is applied for the fiducial acceptance.
\label{fig:main_result}
}
\end{figure}

Figure~\ref{fig:main_result} summarises the
results of the analysis, namely the measurement of the
relative rate of the coincident \rhoz\ production,
which is defined as the ratio of the number of
\gamgammumu\ events with a coincident \rhoz\ ($N_{\gamma \gamma \rightarrow \mu\mu}^{\rhoz}$), to the number
of exclusive \gamgammumu\ events ($N_{\gamma \gamma \rightarrow \mu\mu}^{\mathrm{excl}}$).
The coincidence rates are measured in the fiducial acceptance of the
ATLAS detector, namely for $|\eta|<2.5$ and $\pt>100$~\MeV\ for the
pions from the \rhoz\ decay.
An analysis using the \starlight\ MC generator shows that
approximately 76\% of \rhoz\ mesons with $|y|<2.5$, satisfy this requirement.
The results shown in Figure~\ref{fig:main_result} are fully corrected
to account for the previously discussed reconstruction inefficiencies,
but not corrected for the fiducial acceptance.
The results are shown for the ZDC selections of
\OnOn, \OnXn, \OnYn, \YnYn\ and the combined \XnXn\ selection.
The relative rate for coincident \rhoz\ production is $\sim 5\times10^{-3}$ in the \OnOn\ selection and
increases monotonically with the following change of the selection: $\OnOn\rightarrow\OnYn\rightarrow\YnYn$.
This can be understood as a consequence of increasing neutron activity selecting,
on average, smaller impact parameters and correspondingly a modified photon spectra,
where the probability of coincident production is enhanced.
The maximum of the relative rate is reached for the
\YnYn\ selection with the value of approximately $30\times10^{-3}$.
For the combined \XnXn\ ZDC interval, the relative rate for coincident \rhoz\ production
is $(9.3\,\pm0.4\,\mathrm{(stat.)}\,\pm0.2\,\mathrm{(syst.)})\times10^{-3}$,
which is the same order of magnitude as suggested in Ref.~\cite{Klein:1999qj}.
For each of the ZDC selections, results are also shown
for selections on dimuon invariant mass, \mmumu,
dimuon rapidity, \ymumu, and energy of the leading photon, \egammalead,
shown in the left, middle, and right panels of
Figure~\ref{fig:main_result}, respectively.
The photon energies of the \gamgammumu\ process are not measured directly,
but can be obtained from the dimuon kinematics as
$E_{\gamma}=\frac{\mmumu}{2}e^{\pm \ymumu}$.
The photon energy distributions for the exclusive \gamgammumu\ events
are included in the Appendix.
The $\mmumu<15~\GeV$ and $\egammalead<16~\GeV$ selections
have lower thresholds on \mmumu\ and \egammalead\ of approximately $8~\GeV$ and $4~\GeV$, respectively,
that are introduced by the \pt\ and \ptbar\
requirements imposed on the two muons.
In each panel of Figure~\ref{fig:main_result}, the results are compared
with the nominal case where no selections on kinematic quantities are applied.
The coincidence rate generally increases with \mmumu,
though the large statistical uncertainties in the measurements prohibit observing
any clear trends in the \OnYn\ and \YnYn\ intervals.
This increase in the relative rate with increasing \mmumu\ is reflective of the fact that
a higher \mmumu\ on average corresponds to a smaller
impact parameter of the collision.
For the \OnOn, \OnXn, and \XnXn\ ZDC selections, coincidence rates are
systematically larger when the tagging dimuon is at mid-rapidity.
No clear systematic dependence is observed on the energy of
the leading photon across the different ZDC selections.

While previous measurements have established \rhoz\ production
rates in intervals of forward neutron activity~\cite{STAR:2002caw,STAR:2007elq,STAR:2011wtm,STAR:2017enh,ALICE:2015nbw,ALICE:2020ugp},
the present measurements additionally exploit the presence of a
coincident \gamgammumu\ pair and selections on the
\gamgammumu\ kinematics.
As a result, the various combinations of neutron activity and
dimuon kinematics shown in Figure~\ref{fig:main_result}
are expected to provide significantly enhanced constraints
on the impact-parameter dependence of \rhoz\ production.

\section{Conclusion\label{sec:conclusion}}
In summary, ATLAS measurements are presented of the
relative rate of coincident production of \gamgammumu\ and \rhopipi, including direct \pipi\ production,
in \PbPb\ UPC events.
The measurements are performed using data recorded
at centre-of-mass energies of 5.02~\TeV\ and 5.36~\TeV.
Events are tagged by the presence of a \gamgammumu\ process
with two identified muons.
In these tagged events, those with two additional tracks
are used to identify \rhoz\ candidates.
Template fits to the pair mass
of the \rhoz\ candidates are used to estimate the background
contamination in these tagged events.
The  rate of coincident \rhoz\ production relative to exclusive \gamgammumu\
is found to increase with increasing event activity measured by ZDC as well
as with the mass of the tagging dimuon system.
For the inclusive ZDC and \mmumu\ interval, the relative rate
for the coincident \rhoz\ production is found to be
$(9.3\,\pm0.4\,\mathrm{(stat.)}\,\pm0.2\,\mathrm{(syst.)})\times10^{-3}$.
These measurements provide the first quantitative observation of coincident
photon-induced processes in UPC events and their dependence on
observables sensitive to the impact-parameter.
They also provide input for future tests of the factorized treatment of
independent photon-induced interactions, and for the development of
theoretical calculations and Monte Carlo descriptions of multi-process
UPC events, with potential sensitivity to higher-order electromagnetic
effects in the strong-field regime.


%

%
%
%

%
%
%
%
%
\section*{Acknowledgements}
%

%

%
%

%
%

We thank CERN for the very successful operation of the LHC and its injectors, as well as the support staff at
CERN and at our institutions worldwide without whom ATLAS could not be operated efficiently.

The crucial computing support from all WLCG partners is acknowledged gratefully, in particular from CERN, the ATLAS Tier-1 facilities at TRIUMF/SFU (Canada), NDGF (Denmark, Norway, Sweden), CC-IN2P3 (France), KIT/GridKA (Germany), INFN-CNAF (Italy), NL-T1 (Netherlands), PIC (Spain), RAL (UK) and BNL (USA), the Tier-2 facilities worldwide and large non-WLCG resource providers. Major contributors of computing resources are listed in Ref.~\cite{ATL-SOFT-PUB-2026-001}.

We gratefully acknowledge the support of ANPCyT, Argentina; YerPhI, Armenia; ARC, Australia; BMWFW and FWF, Austria; ANAS, Azerbaijan; CNPq and FAPESP, Brazil; NSERC, NRC and CFI, Canada; CERN; ANID, Chile; CAS, MOST and NSFC, China; Minciencias, Colombia; MEYS CR, Czech Republic; DNRF and DNSRC, Denmark; IN2P3-CNRS and CEA-DRF/IRFU, France; SRNSFG, Georgia; BMFTR, HGF and MPG, Germany; GSRI, Greece; RGC and Hong Kong SAR, China; ICHEP and Academy of Sciences and Humanities, Israel; INFN, Italy; MEXT and JSPS, Japan; CNRST, Morocco; NWO, Netherlands; RCN, Norway; MNiSW, Poland; FCT, Portugal; MNE/IFA, Romania; MSTDI, Serbia; MSSR, Slovakia; ARIS and MVZI, Slovenia; DSI/NRF, South Africa; MICIU/AEI, Spain; SRC and Wallenberg Foundation, Sweden; SERI, SNSF and Cantons of Bern and Geneva, Switzerland; NSTC, Taipei; TENMAK, T\"urkiye; STFC/UKRI, United Kingdom; DOE and NSF, United States of America.

Individual groups and members have received support from BCKDF, CANARIE, CRC and DRAC, Canada; CERN-CZ, FORTE and PRIMUS, Czech Republic; COST, ERC, ERDF, Horizon 2020, ICSC-NextGenerationEU and Marie Sk{\l}odowska-Curie Actions, European Union; Investissements d'Avenir Labex, Investissements d'Avenir Idex and ANR, France; DFG and AvH Foundation, Germany; Herakleitos, Thales and Aristeia programmes co-financed by EU-ESF and the Greek NSRF, Greece; BSF-NSF and MINERVA, Israel; NCN and NAWA, Poland; La Caixa Banking Foundation, CERCA and AGAUR programs from Generalitat de Catalunya and PROMETEO and GenT Programmes Generalitat Valenciana, Spain; G\"{o}ran Gustafssons Stiftelse, Sweden; The Royal Society and Leverhulme Trust, United Kingdom; Eric and Wendy Schmidt Fund for Strategic Innovation, United States of America.

In addition, individual members wish to acknowledge support from Armenia: Yerevan Physics Institute (FAPERJ); CERN: European Organization for Nuclear Research (CERN DOCT); Chile: Agencia Nacional de Investigaci\'on y Desarrollo (ANID FONDECYT reg. 1230987, FONDECYT 1230812, FONDECYT 1240864, Fondecyt 3240661, Fondecyt Regular 1240721); China: Chinese Ministry of Science and Technology (MOST-2023YFA1605700, MOST-2023YFA1609300), National Natural Science Foundation of China (NSFC - 12175119, NSFC 12275265); Czech Republic: Czech Science Foundation (GACR - 24-11373S), Ministry of Education Youth and Sports (ERC-CZ-LL2327, FORTE CZ.02.01.01/00/22\_008/0004632), PRIMUS Research Programme (PRIMUS/21/SCI/017); EU: H2020 European Research Council (ERC - 101002463); European Union: European Research Council (BARD No. 101116429, ERC - 948254, ERC 101089007), European Regional Development Fund (HE COFUND GA No.101081355, ERDF), Horizon 2020 Framework Programme (MUCCA - CHIST-ERA-19-XAI-00), European Union, Future Artificial Intelligence Research (FAIR-NextGenerationEU PE00000013), Horizon 2020 (EuroHPC - EHPC-DEV-2024D11-051), Italian Center for High Performance Computing, Big Data and Quantum Computing (ICSC, NextGenerationEU), Marie Sklodowska-Curie Actions (ERC Horizon-2020, GA 956086, GAP-101168829); France: Agence Nationale de la Recherche (ANR-21-CE31-0013, ANR-21-CE31-0022, ANR-22-EDIR-0002, ANR-24-CE31-0504-01); Germany: Baden-Württemberg Stiftung (BW Stiftung-Postdoc Eliteprogramme), Deutsche Forschungsgemeinschaft (DFG - 469666862, DFG - CR 312/5-2); China: Research Grants Council (GRF); Italy: Istituto Nazionale di Fisica Nucleare (ICSC, NextGenerationEU), Ministero dell'Università e della Ricerca (NextGenEU 153D23001490006 M4C2.1.1, NextGenEU I53D23000820006 M4C2.1.1, NextGenEU I53D23001490006 M4C2.1.1, SOE2024\_0000023); Japan: Japan Society for the Promotion of Science (JSPS KAKENHI JP22H01227, JSPS KAKENHI JP22H04944, JSPS KAKENHI JP22KK0227, JSPS KAKENHI JP24K23939); Norway: Research Council of Norway (RCN-314472); Poland: Ministry of Science and Higher Education (IDUB AGH, POB8, D4 no 9722), Polish National Science Centre (NCN 2021/42/E/ST2/00350, NCN OPUS 2023/51/B/ST2/02507, NCN OPUS nr 2022/47/B/ST2/03059, NCN UMO-2019/34/E/ST2/00393, UMO-2022/47/O/ST2/00148, UMO-2023/49/B/ST2/04085, UMO-2023/51/B/ST2/00920, UMO-2024/53/N/ST2/00869); Portugal: Foundation for Science and Technology (FCT); Spain: Agència de Gestió d'Ajuts Universitaris i de Recerca. (AGAUR - 2023 BP 00141), Generalitat Valenciana (ASFAE/2022/008), Ministry of Science and Innovation (MCIN \& NextGenEU PCI2022-135018-2, MICIN \& FEDER PID2021-125273NB, RYC2019-028510-I, RYC2020-030254-I, RYC2021-031273-I, RYC2022-038164-I), Ministerio de Ciencia, Innovación y Universidades/Agencia Estatal de Investigaci\'on (EU NextGenerationEU (PRTR-C17.I1), PID2022-142604OB-C22); Sweden: Carl Trygger Foundation (Carl Trygger Foundation CTS 22:2312), Swedish Research Council (Swedish Research Council 2023-04654, VR 2021-03651, VR 2022-03845, VR 2022-04683, VR 2023-03403, VR 2024-05451), Knut and Alice Wallenberg Foundation (KAW 2018.0458, KAW 2022.0358, KAW 2023.0366); Switzerland: Swiss National Science Foundation (SNSF - PCEFP2\_194658); United Kingdom: Leverhulme Trust (Leverhulme Trust RPG-2020-004), Royal Society (NIF-R1-231091); United States of America: U.S. Department of Energy (ECA DE-AC02-76SF00515), Neubauer Family Foundation.

%
%


\section*{Data availability}
\printatlasdatapolicy

\clearpage
\appendix
\part*{Appendix}
\label{sec:appendix}
%

\begin{figure}[h]
\centering
\includegraphics[width=0.99\textwidth]{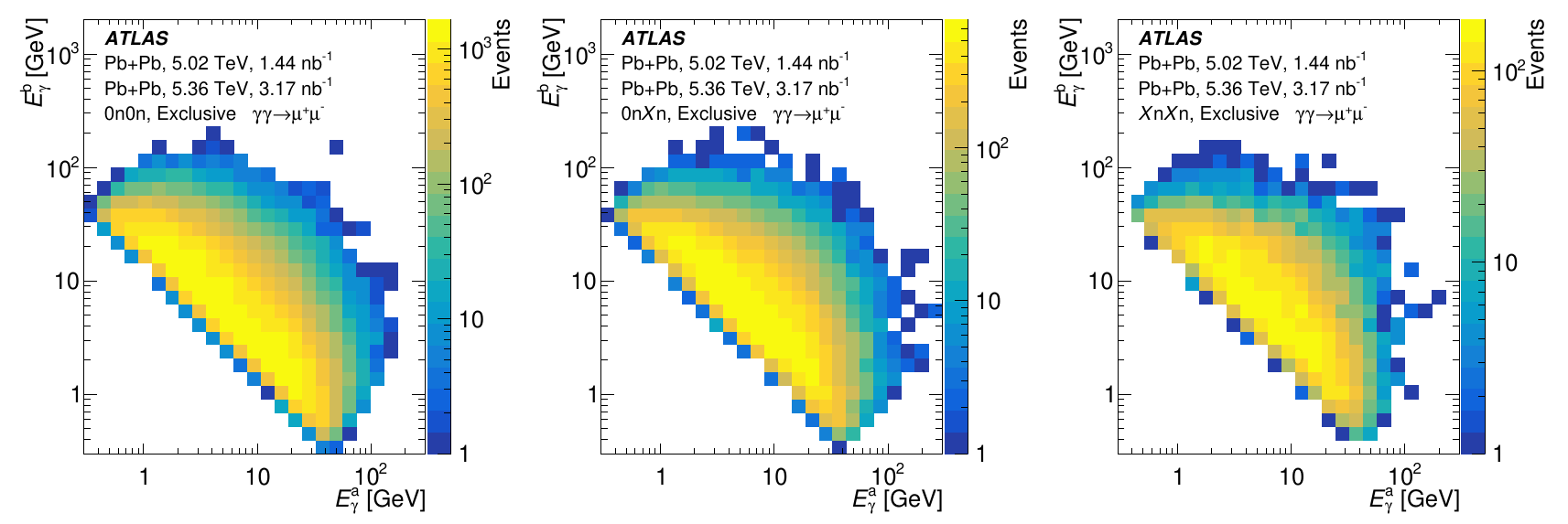}
\caption{
Distributions of the energies of the two photons
in the exclusive \gamgammumu\ events.
The left, centre and right panels correspond to the
\OnOn, \OnYn\ and \YnYn\ ZDC selections, respectively.
\label{fig:DimuonEgammaDist}
}
\end{figure}

\begin{figure}[h]
\centering
\includegraphics[width=0.99\textwidth]{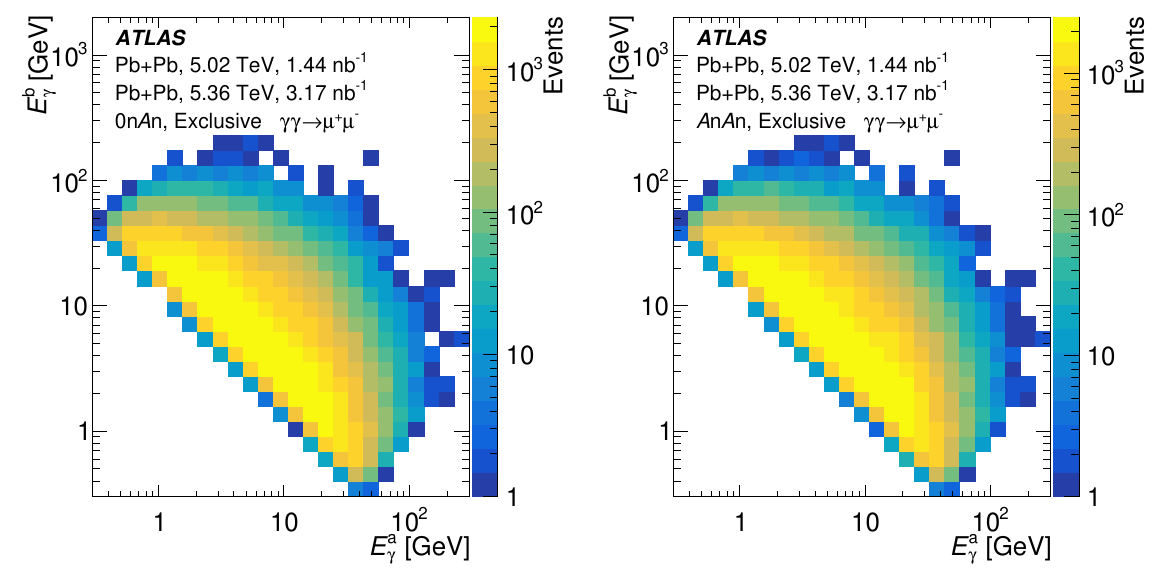}
\caption{
Distributions of the energies of the two photons
in the exclusive \gamgammumu\ events.
The left and right panels correspond to the
\OnXn\ and \XnXn\ ZDC selections, respectively.
\label{fig:DimuonEgammaDist2}
}
\end{figure}

Figures~\ref{fig:DimuonEgammaDist}--\ref{fig:DimuonEgammaDist2}
show the correlation between the energies of the two photons
in the exclusive \gamgammumu\ events.
From these distributions, the dimuon mass, rapidity and the
distributions for the leading/sub-leading photon energy can be obtained
via the relation $E_{\gamma}=\frac{\mmumu}{2}e^{\pm\ymumu}$.

\begin{figure}[h]
\centering
\includegraphics[width=1.0\textwidth]{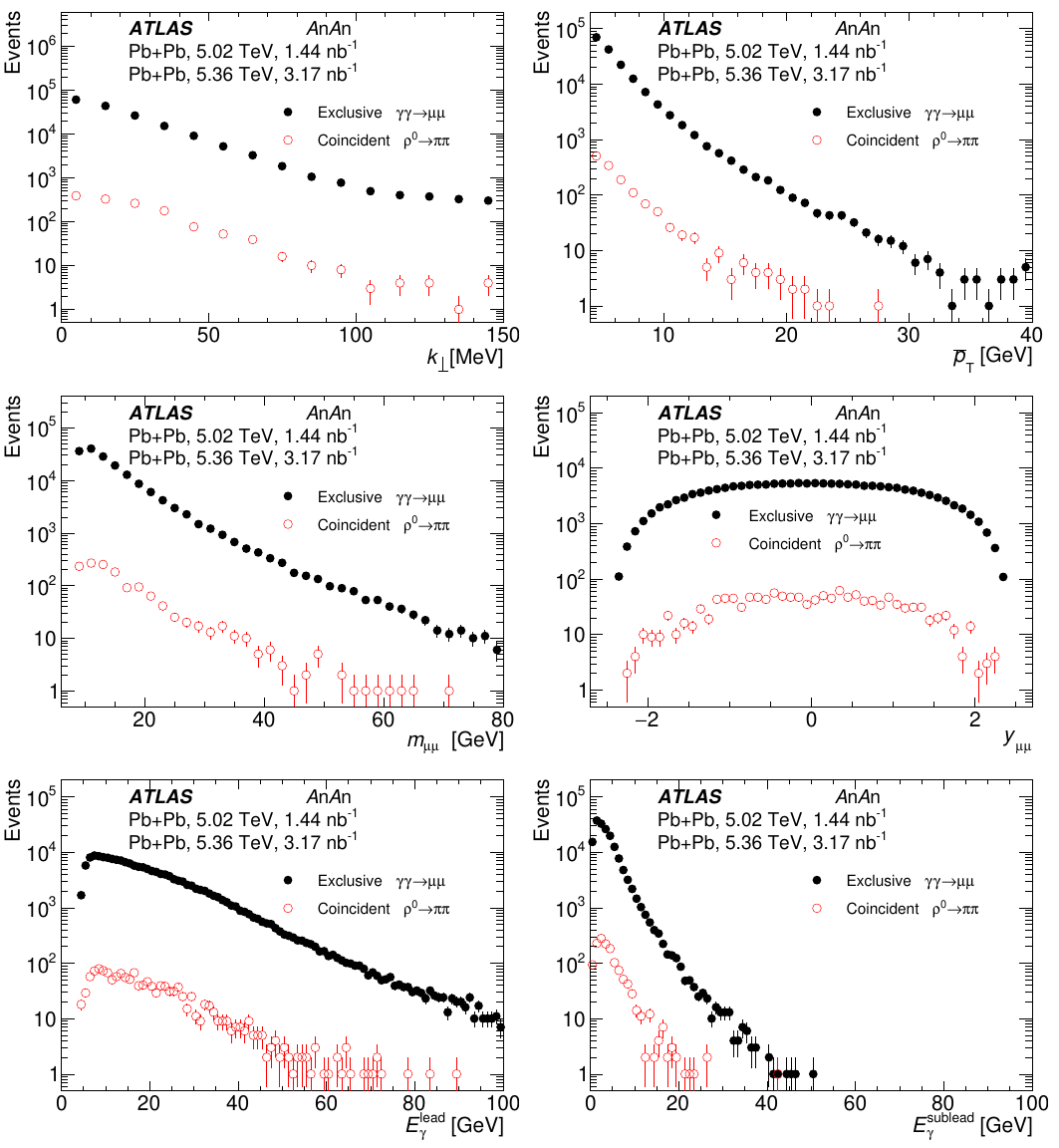}
\caption{
Kinematic distributions of the dimuons in exclusive \gamgammumu\ events and
in events with a coincident \rhoz.
From top left to bottom right the panels correspond to
\kperp, \ptbar, \mmumu, \ymumu, energy of the higher energy photon
and energy of the lower energy photon going into the \gamgammumu\ process.
The distributions are shown for the \XnXn\ ZDC selection.
\label{fig:dimuon_features}
}
\end{figure}

Figure~\ref{fig:dimuon_features} compares the kinematic distributions
of the dimuons in exclusive \gamgammumu\ events with those
in events with a coincident \rhoz\ candidate present.
In general, the dimuon kinematics are qualitatively similar between
the two sets of events.

\FloatBarrier
\printbibliography

@article{Bertulani:1987tz,
    author = "Bertulani, Carlos A. and Baur, Gerhard",
    title = "{Electromagnetic Processes in Relativistic Heavy Ion Collisions}",
    reportNumber = "JUL-2163",
    doi = "10.1016/0370-1573(88)90142-1",
    journal = "Phys. Rept.",
    volume = "163",
    pages = "299",
    year = "1988"
}

@article{STAR:2019wlg,
    author = "Adam, Jaroslav and others",
    collaboration = "STAR",
    title = "{Measurement of $e^+e^-$ Momentum and Angular Distributions from Linearly Polarized Photon Collisions}",
    eprint = "1910.12400",
    archivePrefix = "arXiv",
    primaryClass = "nucl-ex",
    doi = "10.1103/PhysRevLett.127.052302",
    journal = "Phys. Rev. Lett.",
    volume = "127",
    number = "5",
    pages = "052302",
    year = "2021"
}

@article{Klein:2020fmr,
    author = "Klein, Spencer and Steinberg, Peter",
    title = "{Photonuclear and Two-photon Interactions at High-Energy Nuclear Colliders}",
    eprint = "2005.01872",
    archivePrefix = "arXiv",
    primaryClass = "nucl-ex",
    doi = "10.1146/annurev-nucl-030320-033923",
    journal = "Ann. Rev. Nucl. Part. Sci.",
    volume = "70",
    pages = "323--354",
    year = "2020"
}

@article{LHCb:2025fzk,
    author = "{LHCb Collaboration}",
    title = "{Coherent photoproduction of {\ensuremath{\rho}}$^{0}$, {\ensuremath{\omega}} and excited vector mesons in ultraperipheral PbPb collisions}",
    eprint = "2506.06250",
    archivePrefix = "arXiv",
    primaryClass = "nucl-ex",
    reportNumber = "CERN-EP-2025-106, LHCb-PAPER-2024-042",
    doi = "10.1007/JHEP11(2025)103",
    journal = "JHEP",
    volume = "11",
    pages = "103",
    year = "2025"
}

@article{Hencken:2006ir,
    author = "Hencken, K. and Kuraev, E. A. and Serbo, V.",
    title = "{Exclusive and inclusive muon pair production in collisions of relativistic nuclei}",
    eprint = "hep-ph/0606069",
    archivePrefix = "arXiv",
    doi = "10.1103/PhysRevC.75.034903",
    journal = "Phys. Rev. C",
    volume = "75",
    pages = "034903",
    year = "2007"
}

@article{Vermaseren:1982cz,
    author = "Vermaseren, J. A. M.",
    title = "{Two Photon Processes at Very High-Energies}",
    reportNumber = "NIKHEF-H/82-15",
    doi = "10.1016/0550-3213(83)90336-X",
    journal = "Nucl. Phys. B",
    volume = "229",                                                                                                                                                                            
    pages = "347--371",                                                                                                                                                                        
    year = "1983"                                                                                                                                                                              
}

@article{RHIC, 
  author  = "Hahn,  H. and others", 
  title   = "{The RHIC design overview}", 
  doi     = "10.1016/S0168-9002(02)01938-1",
  journal = "Nucl. Instrum. Meth. A", 
  volume  = "499", 
  pages   = "245--263", 
  year    = "2003"
}

@article{Geant4PhysicsLists2,
    author = "Allison, J. and others",
    title = "{Recent developments in \textsc{Geant4}}",
    reportNumber = "FERMILAB-PUB-16-447-CD",
    doi = "10.1016/j.nima.2016.06.125",
    journal = "Nucl. Instrum. Meth. A",
    volume = "835",
    pages = "186--225",
    year = "2016"
}

@article{Baltz:2009jk,
    author = "Baltz, Anthony J. and Gorbunov, Yuri and Klein, Spencer R. and Nystrand, Joakim",
    title = "{Two-Photon Interactions with Nuclear Breakup in Relativistic Heavy Ion Collisions}",
    eprint = "0907.1214",
    archivePrefix = "arXiv",
    primaryClass = "nucl-ex",
    doi = "10.1103/PhysRevC.80.044902",
    journal = "Phys. Rev. C",
    volume = "80",
    pages = "044902",
    year = "2009"
}

@article{Bertulani:2005ru,
    author = "Bertulani, Carlos A. and Klein, Spencer R. and Nystrand, Joakim",
    title = "{Physics of ultra-peripheral nuclear collisions}",
    eprint = "nucl-ex/0502005",
    archivePrefix = "arXiv",
    doi = "10.1146/annurev.nucl.55.090704.151526",
    journal = "Ann. Rev. Nucl. Part. Sci.",
    volume = "55",
    pages = "271--310",
    year = "2005"
}

@article{vanHameren:2017krz,
    author = "van Hameren, Andreas and K\l{}usek-Gawenda, Mariola and Szczurek, Antoni",
    title = "{Single- and double-scattering production of four muons in ultraperipheral $PbPb$ collisions at the Large Hadron Collider}",
    eprint = "1708.07742",
    archivePrefix = "arXiv",
    primaryClass = "hep-ph",
    doi = "10.1016/j.physletb.2017.11.029",
    journal = "Phys. Lett. B",
    volume = "776",
    pages = "84--90",
    year = "2018"
}

@article{Azevedo:2023vsz,
    author = "Azevedo, Celsina N. and Gon\c{c}alves, Victor P. and Moreira, Bruno D.",
    title = "{Double particle production in ultraperipheral PbPb collisions at the Large Hadron Collider and Future Circular Collider}",
    eprint = "2306.05519",
    archivePrefix = "arXiv",
    primaryClass = "hep-ph",
    doi = "10.1140/epja/s10050-023-01114-4",
    journal = "Eur. Phys. J. A",
    volume = "59",
    number = "8",
    pages = "193",
    year = "2023"
}

@article{STAR:2002caw,
    author = "Adler, C. and others",
    collaboration = "STAR",
    title = "{Coherent $\rho^{0}$ production in ultraperipheral heavy ion collisions}",
    eprint = "nucl-ex/0206004",
    archivePrefix = "arXiv",
    doi = "10.1103/PhysRevLett.89.272302",
    journal = "Phys. Rev. Lett.",
    volume = "89",
    pages = "272302",
    year = "2002"
}

@article{STAR:2007elq,
    author = "Abelev, B. I. and others",
    collaboration = "STAR",
    title = "{$\rho^0$ photoproduction in ultraperipheral relativistic heavy ion collisions at $\sqrt{s_{NN}}$ = 200 GeV}",
    eprint = "0712.3320",
    archivePrefix = "arXiv",
    primaryClass = "nucl-ex",
    doi = "10.1103/PhysRevC.77.034910",
    journal = "Phys. Rev. C",
    volume = "77",
    pages = "034910",
    year = "2008"
}

@article{STAR:2011wtm,
    author = "Agakishiev, G. and others",
    collaboration = "STAR",
    title = "{$\rho^{0}$ Photoproduction in AuAu Collisions at $\sqrt{s_{NN}}$=62.4 GeV with STAR}",
    eprint = "1107.4630",
    archivePrefix = "arXiv",
    primaryClass = "nucl-ex",
    doi = "10.1103/PhysRevC.85.014910",
    journal = "Phys. Rev. C",
    volume = "85",
    pages = "014910",
    year = "2012"
}

@article{STAR:2017enh,
    author = "Adamczyk, L. and others",
    collaboration = "STAR",
    title = "{Coherent diffractive photoproduction of $\rho^{0}$ mesons on gold nuclei at 200 GeV/nucleon-pair at the Relativistic Heavy Ion Collider}",
    eprint = "1702.07705",
    archivePrefix = "arXiv",
    primaryClass = "nucl-ex",
    doi = "10.1103/PhysRevC.96.054904",
    journal = "Phys. Rev. C",
    volume = "96",
    number = "5",
    pages = "054904",
    year = "2017"
}

@article{Klein:1999qj,
    author = "Klein, Spencer and Nystrand, Joakim",
    title = "{Exclusive vector meson production in relativistic heavy ion collisions}",
    eprint = "hep-ph/9902259",
    archivePrefix = "arXiv",
    reportNumber = "LBNL-42768, LBL-42768",
    doi = "10.1103/PhysRevC.60.014903",
    journal = "Phys. Rev. C",
    volume = "60",
    pages = "014903",
    year = "1999"
}

@Article{Agostinelli:2002hh,
%     author    = "{GEANT4 Collaboration, S.~Agostinelli et al.}",
%     title     = "{GEANT4: A simulation toolkit}",
%     journal   = "Nucl. Instrum. Meth.",
%     volume    = "A~506",
%     year      = "2003",
%     pages     = "250-303",
%doi            = "10.1016/S0168-9002(03)01368-8",
%}

@article{Klein:2016yzr,
      author         = "Klein, Spencer R. and Nystrand, Joakim and Seger, Janet
                        and Gorbunov, Yuri and Butterworth, Joey",
      title          = "{STARlight: A Monte Carlo simulation program for
                        ultra-peripheral collisions of relativistic ions}",
      journal        = "Comput. Phys. Commun.",
      volume         = "212",
      year           = "2017",
      pages          = "258-268",
      doi            = "10.1016/j.cpc.2016.10.016",
      eprint         = "1607.03838",
      archivePrefix  = "arXiv",
      primaryClass   = "hep-ph",
      SLACcitation   = "%%CITATION = ARXIV:1607.03838;%%"
}

@article{Baur:2001jj,
    author = "Baur, Gerhard and Hencken, Kai and Trautmann, Dirk and Sadovsky, Serguei and Kharlov, Yuri",
    title = "{Coherent $\gamma\gamma$ and $\gamma$A interactions in very peripheral collisions at relativistic ion colliders}",
    eprint = "hep-ph/0112211",
    archivePrefix = "arXiv",
    doi = "10.1016/S0370-1573(01)00101-6",
    journal = "Phys. Rept.",
    volume = "364",
    pages = "359--450",
    year = "2002"
}

@article{Harland-Lang:2021ysd,
    author = "Harland-Lang, L. A. and Khoze, V. A. and Ryskin, M. G.",
    title = "{Elastic photon-initiated production at the LHC: the role of hadron-hadron interactions}",
    eprint = "2104.13392",
    archivePrefix = "arXiv",
    primaryClass = "hep-ph",
    reportNumber = "IPPP/20/97",
    doi = "10.21468/SciPostPhys.11.3.064",
    journal = "SciPost Phys.",
    volume = "11",
    pages = "064",
    year = "2021"
}

@Article{Evans:2008zzb,
      author         = "Evans, Lyndon and Bryant, Philip",
      title          = "{LHC Machine}",
      journal        = "JINST",
      volume         = "3",
      pages          = "S08001",
      doi            = "10.1088/1748-0221/3/08/S08001",
      year           = "2008",
      SLACcitation   = "%%CITATION = JINST,3,S08001;%%",
}

@Article{PERF-2007-01,
    author         = "{ATLAS Collaboration}",
    title          = "{The ATLAS Experiment at the CERN Large Hadron Collider}",
    journal        = "JINST",
    volume         = "3",
    year           = "2008",
    pages          = "S08003",
    doi            = "10.1088/1748-0221/3/08/S08003",
    primaryClass   = "hep-ex",
}

@Article{SOFT-2010-01,
    author         = "{ATLAS Collaboration}",
    title          = "{The ATLAS Simulation Infrastructure}",
    journal        = "Eur. Phys. J. C",
    volume         = "70",
    year           = "2010",
    pages          = "823",
    doi            = "10.1140/epjc/s10052-010-1429-9",
    eprint         = "1005.4568",
    archivePrefix  = "arXiv",
    primaryClass   = "physics.ins-det",
}

@Article{PERF-2015-07,
    author         = "{ATLAS Collaboration}",
    title          = "{Study of the material of the ATLAS inner detector for Run~2 of the LHC}",
    journal        = "JINST",
    volume         = "12",
    year           = "2017",
    pages          = "P12009",
    doi            = "10.1088/1748-0221/12/12/P12009",
    reportNumber   = "CERN-EP-2017-081",
    eprint         = "1707.02826",
    archivePrefix  = "arXiv",
    primaryClass   = "hep-ex",
}

@Article{PERF-2015-08,
    author         = "{ATLAS Collaboration}",
    title          = "{Performance of the ATLAS track reconstruction algorithms in dense environments in LHC Run~2}",
    journal        = "Eur. Phys. J. C",
    volume         = "77",
    year           = "2017",
    pages          = "673",
    doi            = "10.1140/epjc/s10052-017-5225-7",
    reportNumber   = "CERN-EP-2017-045",
    eprint         = "1704.07983",
    archivePrefix  = "arXiv",
    primaryClass   = "hep-ex",
}

@Article{PERF-2015-10,
    author         = "{ATLAS Collaboration}",
    title          = "{Muon reconstruction performance of the ATLAS detector in proton--proton collision data at \(\sqrt{s} = 13\,\text{TeV}\)}",
    journal        = "Eur. Phys. J. C",
    volume         = "76",
    year           = "2016",
    pages          = "292",
    doi            = "10.1140/epjc/s10052-016-4120-y",
    reportNumber   = "CERN-EP-2016-033",
    eprint         = "1603.05598",
    archivePrefix  = "arXiv",
    primaryClass   = "hep-ex",
}

@Article{HION-2016-02,
    author         = "{ATLAS Collaboration}",
    title          = "{Exclusive dimuon production in ultraperipheral Pb+Pb collisions at \(\sqrt{s_{\text{NN}}} = 5.02\,\text{TeV}\) with ATLAS}",
    journal        = "Phys. Rev. C",
    volume         = "104",
    year           = "2021",
    pages          = "024906",
    doi            = "10.1103/PhysRevC.104.024906",
    reportNumber   = "CERN-EP-2020-138",
    eprint         = "2011.12211",
    archivePrefix  = "arXiv",
    primaryClass   = "nucl-ex",
}

@Article{HION-2016-05,
    author         = "{ATLAS Collaboration}",
    title          = "{Evidence for light-by-light scattering in heavy-ion collisions with the ATLAS detector at the LHC}",
    journal        = "Nature Phys.",
    volume         = "13",
    year           = "2017",
    pages          = "852",
    doi            = "10.1038/nphys4208",
    reportNumber   = "CERN-EP-2016-316",
    eprint         = "1702.01625",
    archivePrefix  = "arXiv",
    primaryClass   = "hep-ex",
}

@Article{TRIG-2016-01,
    author         = "{ATLAS Collaboration}",
    title          = "{Performance of the ATLAS trigger system in 2015}",
    journal        = "Eur. Phys. J. C",
    volume         = "77",
    year           = "2017",
    pages          = "317",
    doi            = "10.1140/epjc/s10052-017-4852-3",
    reportNumber   = "CERN-EP-2016-241",
    eprint         = "1611.09661",
    archivePrefix  = "arXiv",
    primaryClass   = "hep-ex",
}

@Article{DAPR-2018-01,
    author         = "{ATLAS Collaboration}",
    title          = "{ATLAS data quality operations and performance for 2015--2018 data-taking}",
    journal        = "JINST",
    volume         = "15",
    year           = "2020",
    pages          = "P04003",
    doi            = "10.1088/1748-0221/15/04/P04003",
    reportNumber   = "CERN-EP-2019-207",
    eprint         = "1911.04632",
    archivePrefix  = "arXiv",
    primaryClass   = "physics.ins-det",
}

@Article{HION-2018-11,
    author         = "{ATLAS Collaboration}",
    title          = "{Observation of Centrality-Dependent Acoplanarity for Muon Pairs Produced via Two-Photon Scattering in Pb+Pb Collisions at \(\sqrt{s_{\text{NN}}} = 5.02\,\text{TeV}\) with the ATLAS Detector}",
    journal        = "Phys. Rev. Lett.",
    volume         = "121",
    year           = "2018",
    pages          = "212301",
    doi            = "10.1103/PhysRevLett.121.212301",
    reportNumber   = "CERN-EP-2018-100",
    eprint         = "1806.08708",
    archivePrefix  = "arXiv",
    primaryClass   = "nucl-ex",
}

@Article{HION-2018-19,
    author         = "{ATLAS Collaboration}",
    title          = "{Observation of Light-by-Light Scattering in Ultraperipheral Pb+Pb Collisions with the ATLAS Detector}",
    journal        = "Phys. Rev. Lett.",
    volume         = "123",
    year           = "2019",
    pages          = "052001",
    doi            = "10.1103/PhysRevLett.123.052001",
    reportNumber   = "CERN-EP-2019-051",
    eprint         = "1904.03536",
    archivePrefix  = "arXiv",
    primaryClass   = "hep-ex",
}

@Article{TRIG-2018-01,
    author         = "{ATLAS Collaboration}",
    title          = "{Performance of the ATLAS muon triggers in Run~2}",
    journal        = "JINST",
    volume         = "15",
    year           = "2020",
    pages          = "P09015",
    doi            = "10.1088/1748-0221/15/09/p09015",
    reportNumber   = "CERN-EP-2020-031",
    eprint         = "2004.13447",
    archivePrefix  = "arXiv",
    primaryClass   = "physics.ins-det",
}

@Article{GENR-2019-02,
    author         = "{ATLAS Collaboration}",
    title          = "{The ATLAS experiment at the CERN Large Hadron Collider: a description of the detector configuration for Run~3}",
    journal        = "JINST",
    volume         = "19",
    year           = "2024",
    pages          = "P05063",
    doi            = "10.1088/1748-0221/19/05/P05063",
    reportNumber   = "CERN-EP-2022-259",
    eprint         = "2305.16623",
    archivePrefix  = "arXiv",
    primaryClass   = "physics.ins-det",
}

@Article{HION-2019-08,
    author         = "{ATLAS Collaboration}",
    title          = "{Measurement of light-by-light scattering and search for axion-like particles with \(2.2\,\text{nb}^{-1}\) of Pb+Pb data with the ATLAS detector}",
    journal        = "JHEP",
    volume         = "03",
    year           = "2021",
    pages          = "243",
    doi            = "10.1007/JHEP03(2021)243",
    reportNumber   = "CERN-EP-2020-135",
    eprint         = "2008.05355",
    archivePrefix  = "arXiv",
    primaryClass   = "hep-ex",
    related        = "HION-2019-08-err",
    relatedstring  = "Erratum:",
}

@Article{HION-2020-10,
    author         = "{ATLAS Collaboration}",
    title          = "{Measurement of muon pairs produced via \(\gamma\gamma\) scattering in nonultraperipheral Pb+Pb collisions at \(\sqrt{s_{\text{NN}}} = 5.02\,\text{TeV}\) with the ATLAS detector}",
    journal        = "Phys. Rev. C",
    volume         = "107",
    year           = "2023",
    pages          = "054907",
    doi            = "10.1103/PhysRevC.107.054907",
    reportNumber   = "CERN-EP-2022-047",
    eprint         = "2206.12594",
    archivePrefix  = "arXiv",
    primaryClass   = "nucl-ex",
}

@Article{HION-2021-16,
    author         = "{ATLAS Collaboration}",
    title          = "{Exclusive dielectron production in ultraperipheral Pb+Pb collisions at \(\sqrt{s_{\text{NN}}} = 5.02\,\text{TeV}\) with ATLAS}",
    journal        = "JHEP",
    volume         = "06",
    year           = "2023",
    pages          = "182",
    doi            = "10.1007/JHEP06(2023)182",
    reportNumber   = "CERN-EP-2022-130",
    eprint         = "2207.12781",
    archivePrefix  = "arXiv",
    primaryClass   = "nucl-ex",
}

@Article{HION-2022-15,
    author         = "{ATLAS Collaboration}",
    title          = "{Measurement of photonuclear jet production in ultra-peripheral Pb+Pb collisions at \(\sqrt{s_{\text{NN}}} = 5.02\,\text{TeV}\) with the ATLAS detector}",
    journal        = "Phys. Rev. D",
    volume         = "111",
    year           = "2025",
    pages          = "052006",
    doi            = "10.1103/PhysRevD.111.052006",
    reportNumber   = "CERN-EP-2024-221",
    eprint         = "2409.11060",
    archivePrefix  = "arXiv",
    primaryClass   = "nucl-ex",
}

@Article{SOFT-2022-02,
    author         = "{ATLAS Collaboration}",
    title          = "{Software and computing for Run~3 of the ATLAS experiment at the LHC}",
    year           = "2025",
    reportNumber   = "CERN-EP-2024-100",
    eprint         = "2404.06335",
    archivePrefix  = "arXiv",
    primaryClass   = "hep-ex",
    doi            = "10.1140/epjc/s10052-024-13701-w", 
    journal        = "Eur. Phys. J. C", 
    volume         = "85", 
    number         = "3", 
    pages          = "234", 
}

@Article{TRIG-2022-01,
    author         = "{ATLAS Collaboration}",
    title          = "{The ATLAS trigger system for LHC Run~3 and trigger performance in 2022}",
    journal        = "JINST",
    volume         = "19",
    year           = "2024",
    pages          = "P06029",
    doi            = "10.1088/1748-0221/19/06/P06029",
    reportNumber   = "CERN-EP-2023-299",
    eprint         = "2401.06630",
    archivePrefix  = "arXiv",
    primaryClass   = "hep-ex",
}

@Booklet{ATL-SOFT-PUB-2026-001,
    author         = "{ATLAS Collaboration}",
    title          = "{ATLAS Computing Acknowledgements}",
    howpublished   = "{ATL-SOFT-PUB-2026-001}",
    url            = "https://cds.cern.ch/record/2952666",
    year           = 2026
}

@Article{CMS-HIN-12-009,
    author         = "{CMS Collaboration}",
    title          = "{Coherent \(J/\psi\) photoproduction in ultra-peripheral PbPb collisions at \(\sqrt{s_{\text{NN}}} = 2.76\,\text{TeV}\) with the CMS experiment}",
    journal        = "Phys. Lett. B",
    volume         = "772",
    year           = "2017",
    pages          = "489",
    doi            = "10.1016/j.physletb.2017.07.001",
    reportNumber   = "CERN-EP-2016-098",
    eprint         = "1605.06966",
    archivePrefix  = "arXiv",
    primaryClass   = "hep-ex",
}

@Article{CMS-FSQ-16-007,
    author         = "{CMS Collaboration}",
    title          = "{Measurement of exclusive \(\rho(770)^0\) photoproduction in ultraperipheral \(p\)Pb collisions at \(\sqrt{s_{\text{NN}}} = 5.02\,\text{TeV}\)}",
    journal        = "Eur. Phys. J. C",
    volume         = "79",
    year           = "2019",
    pages          = "702",
    doi            = "10.1140/epjc/s10052-019-7202-9",
    reportNumber   = "CERN-EP-2018-285",
    eprint         = "1902.01339",
    archivePrefix  = "arXiv",
    primaryClass   = "hep-ex",
}

@Article{CMS-FSQ-16-012,
    author         = "{CMS Collaboration}",
    title          = "{Evidence for light-by-light scattering and searches for axion-like particles in ultraperipheral PbPb collisions at \(\sqrt{s_{\text{NN}}} = 5.02\,\text{TeV}\)}",
    journal        = "Phys. Lett. B",
    volume         = "797",
    year           = "2019",
    pages          = "134826",
    doi            = "10.1016/j.physletb.2019.134826",
    reportNumber   = "CERN-EP-2018-271",
    eprint         = "1810.04602",
    archivePrefix  = "arXiv",
    primaryClass   = "hep-ex",
}

@Article{CMS-HIN-21-009,
    author         = "{CMS Collaboration}",
    title          = "{Observation of \(\tau\) Lepton Pair Production in Ultraperipheral Pb--Pb Collisions at \(\sqrt{s_{\text{NN}}} = 5.02\,\text{TeV}\)}",
    journal        = "Phys. Rev. Lett.",
    volume         = "131",
    year           = "2023",
    pages          = "151803",
    doi            = "10.1103/PhysRevLett.131.151803",
    reportNumber   = "CERN-EP-2022-098",
    eprint         = "2206.05192",
    archivePrefix  = "arXiv",
    primaryClass   = "nucl-ex",
}

@Article{CMS-HIN-21-015,
    author         = "{CMS Collaboration}",
    title          = "{Measurement of light-by-light scattering and the Breit--Wheeler process, and search for axion-like particles in ultraperipheral PbPb collisions at \(\sqrt{s_{\text{NN}}} = 5.02\,\text{TeV}\)}",
    year           = "2024",
    reportNumber   = "CERN-EP-2024-284",
    eprint         = "2412.15413",
    archivePrefix  = "arXiv",
    primaryClass   = "nucl-ex",
}

@Booklet{ATL-PHYS-PUB-2015-026,
    author         = "{ATLAS Collaboration}",
    title          = "{Vertex Reconstruction Performance of the ATLAS Detector at \(\sqrt{s} = 13~\text{TeV}\)}",
    howpublished   = "{ATL-PHYS-PUB-2015-026}",
    url            = "https://cds.cern.ch/record/2037717",
    year           = "2015",
}

@Booklet{ATL-PHYS-PUB-2019-015,
    author         = "{ATLAS Collaboration}",
    title          = "{Development of ATLAS Primary Vertex Reconstruction for LHC Run 3}",
    howpublished   = "{ATL-PHYS-PUB-2019-015}",
    url            = "https://cds.cern.ch/record/2670380",
    year           = "2019",
}

@Booklet{ATL-PHYS-PUB-2022-033,
    author         = "{ATLAS Collaboration}",
    title          = "{Performance of ATLAS Pixel Detector and Track Reconstruction at the start of Run~3 in LHC Collisions at \(\sqrt{s} = 900\,\text{GeV}\)}",
    howpublished   = "{ATL-PHYS-PUB-2022-033}",
    url            = "https://cds.cern.ch/record/2814766",
    year           = "2022",
}

@article{ALICE:2024ife,
    %author = "Acharya, Shreyasi and others",
    %collaboration = "ALICE",
    author ="{ALICE Collaboration}",
    title = "{Measurement of the impact-parameter dependent azimuthal anisotropy in coherent \ensuremath{\rho^{0}} photoproduction in Pb\textendash{}Pb collisions at $\sqrt{s_\mathrm{{NN}}}=5.02$ TeV}",
    eprint = "2405.14525",
    archivePrefix = "arXiv",
    primaryClass = "nucl-ex",
    reportNumber = "CERN-EP-2024-126",
    doi = "10.1016/j.physletb.2024.139017",
    journal = "Phys. Lett. B",
    volume = "858",
    pages = "139017",
    year = "2024"
}

@article{ALICE:2024kjy,
    %author = "Acharya, Shreyasi and others",
    %collaboration = "ALICE",
    author = "{ALICE Collaboration}",
    title = "{Exclusive four pion photoproduction in ultraperipheral Pb-Pb collisions at $\sqrt{s_{\mathrm{NN}}} = 5.02$ TeV}",
    eprint = "2404.07542",
    archivePrefix = "arXiv",
    primaryClass = "nucl-ex",
    reportNumber = "CERN-EP-2024-104",
    month = "4",
    year = "2024"
}

@article{ALICE:2023jgu,
    %author = "Acharya, Shreyasi and others",
    %collaboration = "ALICE",
    author ="{ALICE Collaboration}",
    title = "{Energy dependence of coherent photonuclear production of J/\ensuremath{\psi} mesons in ultra-peripheral Pb-Pb collisions at $ \sqrt{{\textrm{s}}_{\textrm{NN}}} $ = 5.02 TeV}",
    eprint = "2305.19060",
    archivePrefix = "arXiv",
    primaryClass = "nucl-ex",
    reportNumber = "CERN-EP-2023-100",
    doi = "10.1007/JHEP10(2023)119",
    journal = "JHEP",
    volume = "10",
    pages = "119",
    year = "2023"
}

@article{ALICE:2023mfc,
    author = "{ALICE Collaboration}",
    title = "{Exclusive and dissociative J/\ensuremath{\psi} photoproduction, and exclusive dimuon production, in p-Pb collisions at sNN=8.16\,\,TeV}",
    eprint = "2304.12403",
    archivePrefix = "arXiv",
    primaryClass = "nucl-ex",
    reportNumber = "CERN-EP-2023-059",
    doi = "10.1103/PhysRevD.108.112004",
    journal = "Phys. Rev. D",
    volume = "108",
    number = "11",
    pages = "112004",
    year = "2023"
}

@article{ALICE:2023svb,
    %author = "Acharya, Shreyasi and others",
    %collaboration = "ALICE",
    author ="{ALICE Collaboration}",
    title = "{First polarisation measurement of coherently photoproduced J/$\psi$ in ultra-peripheral Pb$-$Pb collisions at $\sqrt{s_{\mathrm{NN}}}$ = 5.02 TeV}",
    eprint = "2304.10928",
    archivePrefix = "arXiv",
    primaryClass = "nucl-ex",
    reportNumber = "CERN-EP-2023-068",
    month = "4",
    year = "2023"
}

@article{ALICE:2021gpt,
    %author = "Acharya, Shreyasi and others",
    %collaboration = "ALICE",
    author ="{ALICE Collaboration}",
    title = "{Coherent $J/\psi$ and $\psi'$ photoproduction at midrapidity in ultra-peripheral Pb-Pb collisions at $\sqrt{s_{\mathrm{NN}}}~=~5.02$ TeV}",
    eprint = "2101.04577",
    archivePrefix = "arXiv",
    primaryClass = "nucl-ex",
    reportNumber = "CERN-EP-2021-002",
    doi = "10.1140/epjc/s10052-021-09437-6",
    journal = "Eur. Phys. J. C",
    volume = "81",
    number = "8",
    pages = "712",
    year = "2021"
}

@article{ALICE:2021jnv,
    %author = "Acharya, Shreyasi and others",
    %collaboration = "ALICE",
    author ="{ALICE Collaboration}",
    title = "{First measurement of coherent \ensuremath{\rho^{0}} photoproduction in ultra-peripheral Xe\textendash{}Xe collisions at $\sqrt{s_\mathrm{{NN}}}=5.44$ TeV}",
    eprint = "2101.02581",
    archivePrefix = "arXiv",
    primaryClass = "nucl-ex",
    reportNumber = "CERN-EP-2020-252",
    doi = "10.1016/j.physletb.2021.136481",
    journal = "Phys. Lett. B",
    volume = "820",
    pages = "136481",
    year = "2021"
}

@article{ALICE:2020ugp,
    %author = "Acharya, Shreyasi and others",
    %collaboration = "ALICE",
    author ="{ALICE Collaboration}",
    title = "{Coherent photoproduction of $\rho^{0}$ vector mesons in ultra-peripheral Pb-Pb collisions at $ \sqrt{{\mathrm{s}}_{\mathrm{NN}}} $ = 5.02 TeV}",
    eprint = "2002.10897",
    archivePrefix = "arXiv",
    primaryClass = "nucl-ex",
    reportNumber = "CERN-EP-2020-021",
    doi = "10.1007/JHEP06(2020)035",
    journal = "JHEP",
    volume = "06",
    pages = "035",
    year = "2020"
}

@article{ALICE:2019tqa,
    %author = "Acharya, Shreyasi and others",
    %collaboration = "ALICE",
    author ="{ALICE Collaboration}",
    title = "{Coherent J/$\psi$ photoproduction at forward rapidity in ultra-peripheral Pb-Pb collisions at $\sqrt{s_{\mathrm{NN}}}=5.02$ TeV}",
    eprint = "1904.06272",
    archivePrefix = "arXiv",
    primaryClass = "nucl-ex",
    reportNumber = "CERN-EP-2019-069",
    doi = "10.1016/j.physletb.2019.134926",
    journal = "Phys. Lett. B",
    volume = "798",
    pages = "134926",
    year = "2019"
}

@article{ALICE:2018oyo,
    %author = "Acharya, Shreyasi and others",
    %collaboration = "ALICE",
    author ="{ALICE Collaboration}",
    title = "{Energy dependence of exclusive $\mathrm {J}/\psi $ photoproduction off protons in ultra-peripheral p\textendash{}Pb collisions at $\sqrt{s_{\mathrm {\scriptscriptstyle NN}}} = 5.02$ TeV}",
    eprint = "1809.03235",
    archivePrefix = "arXiv",
    primaryClass = "nucl-ex",
    reportNumber = "CERN-EP-2018-236",
    doi = "10.1140/epjc/s10052-019-6816-2",
    journal = "Eur. Phys. J. C",
    volume = "79",
    number = "5",
    pages = "402",
    year = "2019"
}

@article{ALICE:2015nbw,
    %author = "Adam, Jaroslav and others",
    %collaboration = "ALICE",
    author ="{ALICE Collaboration}",
    title = "{Coherent \ensuremath{\rho}$^{0}$ photoproduction in ultra-peripheral Pb-Pb collisions at $ \sqrt{s_{\mathrm{NN}}}=2.76 $ TeV}",
    eprint = "1503.09177",
    archivePrefix = "arXiv",
    primaryClass = "nucl-ex",
    reportNumber = "CERN-PH-EP-2015-082",
    doi = "10.1007/JHEP09(2015)095",
    journal = "JHEP",
    volume = "09",
    pages = "095",
    year = "2015"
}

@article{ALICE:2014eof,
    %author = "Abelev, Betty Bezverkhny and others",
    %collaboration = "ALICE",
    author ="{ALICE Collaboration}",
    title = "{Exclusive $\mathrm{J/}\psi$ photoproduction off protons in ultra-peripheral p-Pb collisions at $\sqrt{s_{\mathrm{NN}}}=5.02$ TeV}",
    eprint = "1406.7819",
    archivePrefix = "arXiv",
    primaryClass = "nucl-ex",
    reportNumber = "CERN-PH-EP-2014-149",
    doi = "10.1103/PhysRevLett.113.232504",
    journal = "Phys. Rev. Lett.",
    volume = "113",
    number = "23",
    pages = "232504",
    year = "2014"
}

@article{ALICE:2013wjo,
    author = "{ALICE Collaboration}",
    title = "{Charmonium and $e^+e^-$ pair photoproduction at mid-rapidity in ultra-peripheral Pb-Pb collisions at $\sqrt{s_{\mathrm{NN}}}$=2.76 TeV}",
    eprint = "1305.1467",
    archivePrefix = "arXiv",
    primaryClass = "nucl-ex",
    reportNumber = "CERN-PH-EP-2013-066",
    doi = "10.1140/epjc/s10052-013-2617-1",
    journal = "Eur. Phys. J. C",
    volume = "73",
    number = "11",
    pages = "2617",
    year = "2013"
}

@article{ALICE:2012yye,
    %author = "Abelev, Betty and others",
    %collaboration = "ALICE",
    author ="{ALICE Collaboration}",
    title = "{Coherent $J/\psi$ photoproduction in ultra-peripheral Pb-Pb collisions at $\sqrt{s_{NN}} = 2.76$ TeV}",
    eprint = "1209.3715",
    archivePrefix = "arXiv",
    primaryClass = "nucl-ex",
    reportNumber = "CERN-PH-EP-2012-270",
    doi = "10.1016/j.physletb.2012.11.059",
    journal = "Phys. Lett. B",
    volume = "718",
    pages = "1273--1283",
    year = "2013"
}
\clearpage
 
\begin{flushleft}
\hypersetup{urlcolor=black}
{\Large The ATLAS Collaboration}

\bigskip

\AtlasOrcid[0000-0002-6665-4934]{G.~Aad}$^\textrm{\scriptsize 104}$,
\AtlasOrcid[0000-0001-7616-1554]{E.~Aakvaag}$^\textrm{\scriptsize 17}$,
\AtlasOrcid[0000-0002-5888-2734]{B.~Abbott}$^\textrm{\scriptsize 122}$,
\AtlasOrcid[0000-0002-0287-5869]{S.~Abdelhameed}$^\textrm{\scriptsize 85b}$,
\AtlasOrcid[0000-0002-1002-1652]{K.~Abeling}$^\textrm{\scriptsize 55}$,
\AtlasOrcid[0000-0001-5763-2760]{N.J.~Abicht}$^\textrm{\scriptsize 49}$,
\AtlasOrcid[0000-0002-8496-9294]{S.H.~Abidi}$^\textrm{\scriptsize 30}$,
\AtlasOrcid[0009-0003-6578-220X]{M.~Aboelela}$^\textrm{\scriptsize 45}$,
\AtlasOrcid[0000-0002-9987-2292]{A.~Aboulhorma}$^\textrm{\scriptsize 36e}$,
\AtlasOrcid[0000-0001-5329-6640]{H.~Abramowicz}$^\textrm{\scriptsize 156}$
% ...
\bigskip
\\

% ...
$^{17}$Department for Physics and Technology, University of Bergen, Bergen; Norway.\\
% ...
$^{30}$Physics Department, Brookhaven National Laboratory, Upton NY; United States of America.\\
% ...
$^{36}$$^{(a)}$Facult\'e des Sciences Ain Chock, Universit\'e Hassan II de Casablanca;$^{(b)}$Facult\'{e} des Sciences, Universit\'{e} Ibn-Tofail, K\'{e}nitra;$^{(c)}$Facult\'e des Sciences Semlalia, Universit\'e Cadi Ayyad, LPHEA-Marrakech;$^{(d)}$LPMR, Facult\'e des Sciences, Universit\'e Mohamed Premier, Oujda;$^{(e)}$Facult\'e des sciences, Universit\'e Mohammed V, Rabat;$^{(f)}$Institute of Applied Physics, Mohammed VI Polytechnic University, Ben Guerir; Morocco.\\
% ...
$^{45}$Physics Department, Southern Methodist University, Dallas TX; United States of America.\\
% ...
$^{49}$Fakult\"{a}t Physik, Technische Universit{\"a}t Dortmund, Dortmund; Germany.\\
% ...
$^{55}$II. Physikalisches Institut, Georg-August-Universit\"{a}t G\"ottingen, G\"ottingen; Germany.\\
% ...
$^{85}$$^{(a)}$Khalifa University of Science and Technology, Abu Dhabi;$^{(b)}$New York University Abu Dhabi, Abu Dhabi;$^{(c)}$United Arab Emirates University, Al Ain;$^{(d)}$University of Sharjah, Sharjah; United Arab Emirates.\\
% ...
$^{104}$CPPM, Aix-Marseille Universit\'e, CNRS/IN2P3, Marseille; France.\\
% ...
$^{122}$Homer L. Dodge Department of Physics and Astronomy, University of Oklahoma, Norman OK; United States of America.\\
% ...
$^{156}$Raymond and Beverly Sackler School of Physics and Astronomy, Tel Aviv University, Tel Aviv; Israel.\\
% ...
$^{a}$ Also at Affiliated with an institute formerly covered by a cooperation agreement with CERN.\\
$^{b}$ Also at An-Najah National University, Nablus; Palestine.\\
$^{c}$ Also at Borough of Manhattan Community College, City University of New York, New York NY; United States of America.\\
$^{d}$ Also at Center for Interdisciplinary Research and Innovation (CIRI-AUTH), Thessaloniki; Greece.\\
$^{e}$ Also at Centre of Physics of the Universities of Minho and Porto (CF-UM-UP); Portugal.\\
$^{f}$ Also at CERN, Geneva; Switzerland.\\
% ...

\end{flushleft}


%

%
%

%

%
%
%
%

\end{document}